\documentclass[]{spie}  

\usepackage{amsmath,amsfonts,amssymb}
\usepackage{graphicx}
\usepackage[colorlinks=true, allcolors=blue]{hyperref}

\title{Three years of harvest with the vector vortex coronagraph in~the thermal infrared}

\author[a]{Olivier Absil}
\author[b,c]{Dimitri Mawet}
\author[d]{Mikael Karlsson}
\author[a]{Brunella Carlomagno}
\author[e,a]{Valentin~Christiaens}
\author[a]{Denis Defr\`ere}
\author[f]{Christian Delacroix}
\author[g]{Bruno Femen\'ia Castell\'a}
\author[d]{Pontus~Forsberg}
\author[h]{Julien Girard}
\author[a]{Carlos A.~G\'omez Gonz\'alez}
\author[a]{Serge Habraken}
\author[i]{Philip~M.~Hinz}
\author[a]{Elsa~Huby}
\author[a]{A\"issa Jolivet}
\author[b]{Keith Matthews}
\author[h]{Julien Milli}
\author[a]{Gilles~Orban~de~Xivry}
\author[j]{Eric~Pantin}
\author[d]{Pierre Piron}
\author[a]{Maddalena Reggiani}
\author[b]{Garreth J.~Ruane}
\author[c]{Eugene Serabyn}
\author[a]{Jean~Surdej}
\author[h]{Konrad~R.~W.~Tristram}
\author[d]{Ernesto~Vargas Catal\'an}
\author[a]{Olivier~Wertz}
\author[g]{Peter~Wizinowich}

\affil[a]{Space sciences, Technologies, and Astrophysics Research (STAR) Institute, Universit\'e~de~Li\`ege, 19c all\'ee du Six Ao\^ut, B-4000 Sart Tilman, Belgium}
\affil[b]{Department of Astronomy, California Institute of Technology, 1200 E. California Blvd, Pasadena, CA 91125, USA}
\affil[c]{Jet Propulsion Laboratory, California Institute of Technology, 4800 Oak Grove Dr., Pasadena, CA 91109, USA}
\affil[d]{Angstr\"om Laboratory, Uppsala University, L\"agerhyddsv\"agen 1, SE-751 21 Uppsala, Sweden}
\affil[e]{Departamento de Astronom\'ia, Universidad de Chile, Casilla 36-D, Santiago, Chile}
\affil[f]{Sibley School of Mechanical and Aerospace Engineering, Cornell University, Ithaca, USA}
\affil[g]{W.\ M.\ Keck Observatory, 65-1120 Mamalahoa Hwy., Kamuela, HI 96743, USA}
\affil[h]{European Southern Observatory, Alonso de C\'ordova 3107, Vitacura, Santiago, Chile}
\affil[i]{Steward Observatory, University of Arizona, 633 N. Cherry Avenue, 85721 Tucson, USA}
\affil[j]{Laboratoire AIM, CEA/DSM -- CNRS -- Univ.\ Paris Diderot, IRFU/SAp, 91191 Gif-sur-Yvette, France}

\authorinfo{Further author information: send correspondence to O.A.\ (absil@astro.ulg.ac.be).}

\begin{document} 
\maketitle

\begin{abstract}
For several years, we have been developing vortex phase masks based on sub-wavelength gratings, known as Annular Groove Phase Masks. Etched onto diamond substrates, these AGPMs are currently designed to be used in the thermal infrared (ranging from 3 to 13~$\mu$m). Our AGPMs were first installed on VLT/NACO and VLT/VISIR in 2012, followed by LBT/LMIRCam in 2013 and Keck/NIRC2 in 2015. In this paper, we review the development, commissioning, on-sky performance, and early scientific results of these new coronagraphic modes and report on the lessons learned. We conclude with perspectives for future developments and applications.
\end{abstract}


\section{INTRODUCTION}
\label{sec:intro}  

Over the past ten years, direct imaging has become one of the main tools for the detection and characterization of planetary systems, including proto-planetary disks, debris disks, and giant extrasolar planets. Direct imaging enables not only spectral characterization of the detected objects, but also to obtain direct information of planetary systems architecture at various ages (including the proto-planetary phase), which is key to our understanding of how planetary systems form and evolve.

In the context of direct planet imaging, the vortex coronagraph\cite{Mawet05,Foo05} has been considered for more than ten years as one of the most promising concepts to reduce the stellar glare\cite{Guyon06}. Yet, vortex coronagraphs have only recently been installed on 10-m class telescopes. Vortex coronagraphs feature a vortex phase mask in their focal plane. The textbook effect of the vortex phase mask is to move the light of an on-axis source outside the geometric image of the input pupil. When followed by a---usually undersized---diaphragm (aka Lyot stop) in a downstream pupil plane to block the diffracted light, the vortex phase mask provides a theoretically perfect starlight cancellation for a clear, circular aperture. One of the possible implementations of the vortex phase mask is the Annular Groove Phase Mask (AGPM)\cite{Mawet05}, made up of a concentric sub-wavelength grating etched onto a diamond substrate\cite{Forsberg13,Delacroix13}. Sub-wavelength gratings produce form birefringence, which can be used to synthesize a helical phase ramp for the two orthogonal polarization states of light separately. By tuning the grating parameters, one can adjust the wavelength-dependence of the birefringent effect induced by the sub-wavelength grating to produce quasi-achromatic behavior across a wide bandwidth\cite{Mawet05b,Delacroix12}. Together with its high throughput and $360^{\circ}$ discovery space, this makes the AGPM an excellent candidate for installation on high-contrast imaging instruments.

Here, we review the design, manufacturing, and testing of these AGPMs (Sect.~\ref{sec:design}). We then report on their installation and commissioning on 10-m class telescopes, including operational aspects  (Sect.~\ref{sec:comm}), and present their measured on-sky performance (Sect.~\ref{sec:perfo}), as well as a few examples of early scientific results  (Sect.~\ref{sec:science}). We conclude with a discussion of the lessons learned from three years of on-sky operations, and of the perspectives for the infrared vortex coronagraph in the coming years  (Sect.~\ref{sec:lessons}).

\section{DESIGN, MANUFACTURING AND LAB TESTING}
\label{sec:design}

Synthetic diamond was chosen as an appropriate material to manufacture AGPMs, thanks to its superb optical, mechanical and thermal properties. Because the period of the sub-wavelength grating scales with the operating wavelength, we have started the development of AGPMs at the longest wavelengths where stellar coronagraphy is considered as a useful technique, i.e., the thermal infrared regime. We focused in particular on the N band (8--13~$\mu$m) and the L band (3.5--4.1~$\mu$m), where the grating period must be of the order of 4~$\mu$m and 1.5~$\mu$m, respectively, to ensure operation in the sub-wavelength regime (i.e., the regime where only the zeroth diffraction order is propagated through the grating).

The design of the sub-wavelength grating is based on rigorous coupled wave analysis (RCWA)\cite{Mawet05}, under the hypothesis that the grating is large enough and has a sufficiently low curvature to be considered straight and infinite. Although this hypothesis breaks down at the very center of the grating, where the curvature becomes large, RCWA is expected to give sufficiently accurate results to enable manufacturing of science-grade AGPMs. The aim of the RCWA modeling is to reach a quasi-achromatic $\pi$ phase shift between the transverse electric (TE) and transverse magnetic (TM) modes of propagation through the grating, i.e., to produce a quasi-achromatic birefringent behavior for the sub-wavelength grating, which then acts locally as an achromatic half wave plate.\cite{Mawet05b} Designs of achromatic half wave plates based on sub-wavelength gratings can be found in previous publications\cite{Delacroix12,Forsberg16,Vargas17}. For example, the optimal parameters for a diamond AGPM covering the L band are a period $\Lambda =1.42~\mu$m, a grating depth $d=5.0~\mu$m, and line width $w_t=0.67~\mu$m at the top of the grating, assuming an angle of 2.45$^{\circ}$ for the sidewalls\cite{Vargas17}.

\begin{figure}[t]
\begin{center}
\begin{tabular}{c} 
\includegraphics[width=8cm]{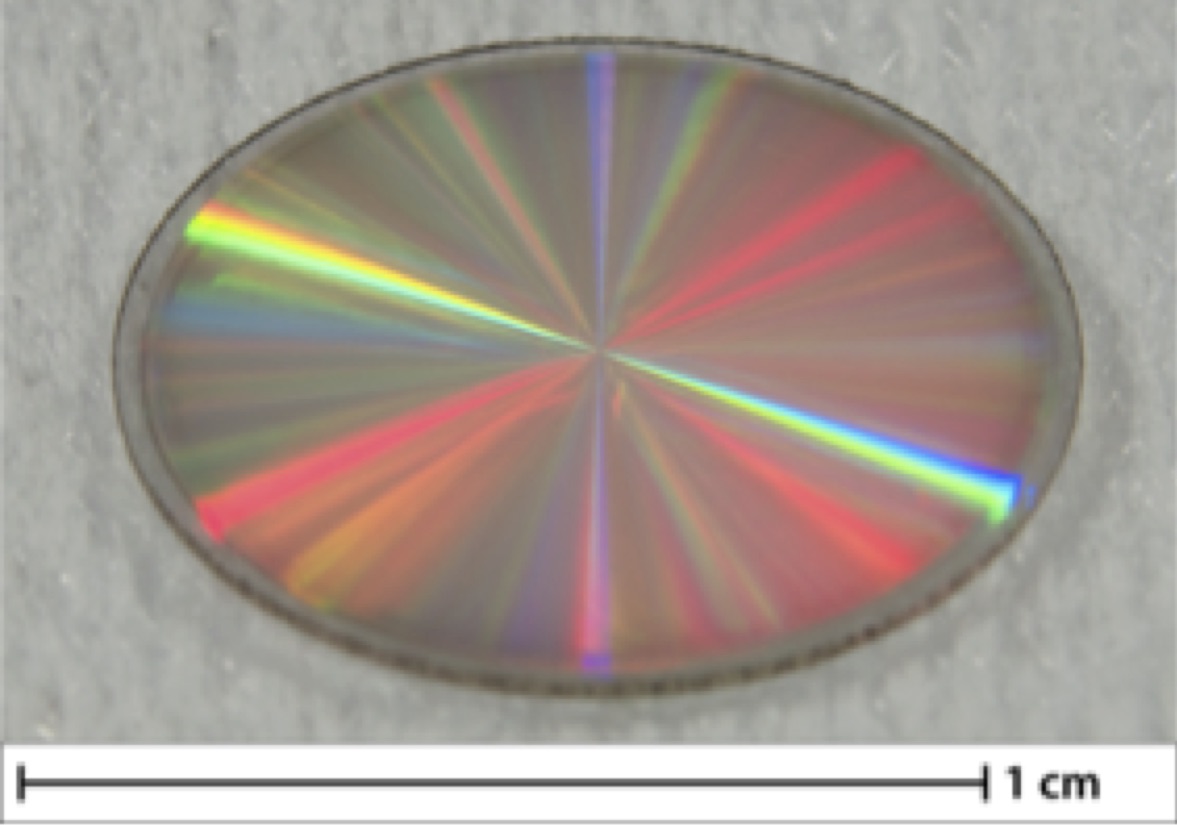} \hspace*{2mm} \includegraphics[width=7.5cm]{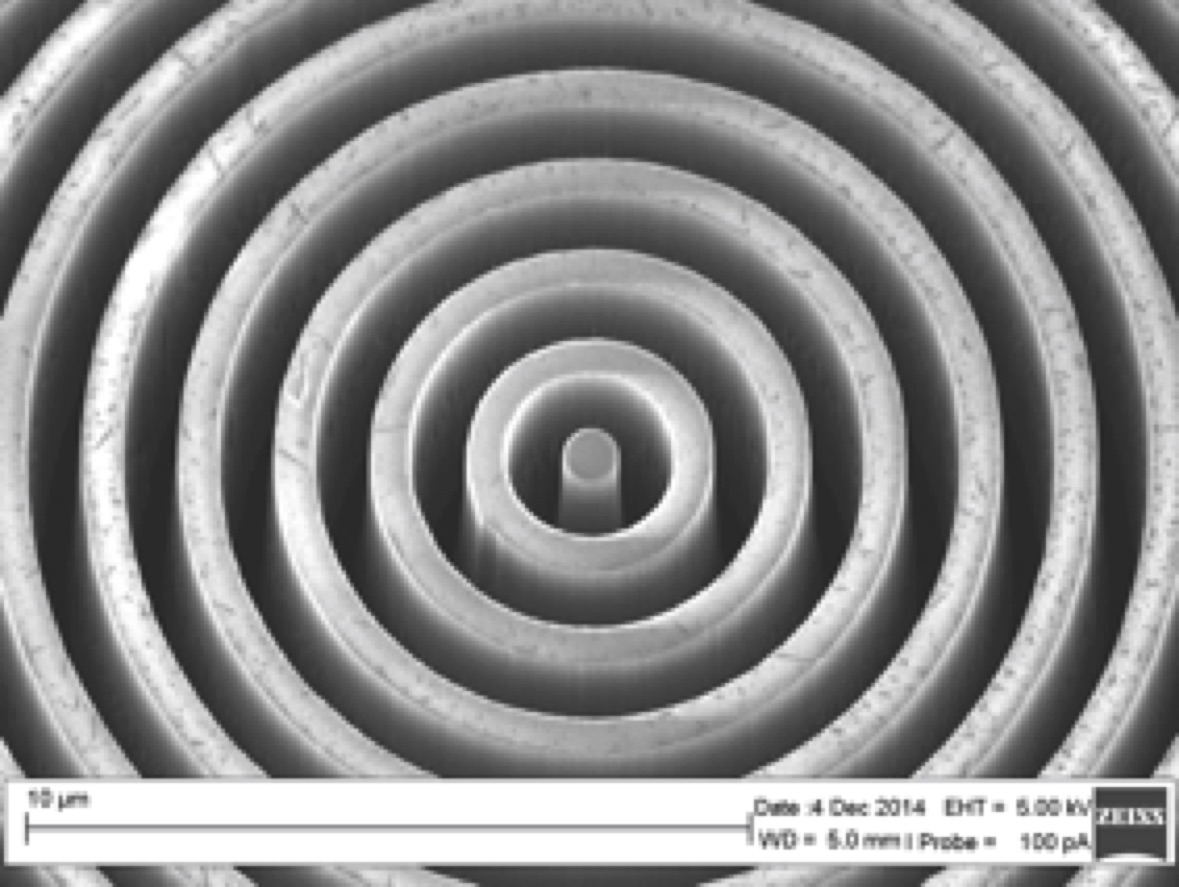}
\end{tabular}
\end{center}
\caption{\label{fig:agpm} \textit{Left.} Photograph of an L-band AGPM, ready for optical evaluation. \textit{Right.} SEM micrograph of the center of an L-band AGPM.}
\end{figure}

Etching high-aspect ratio gratings into diamond has been demonstrated at the Uppsala University in the context of this project\cite{Forsberg13,Vargas16,Forsberg16}. The transfer of the AGPM pattern into a diamond substrate follows a three step process:
\begin{enumerate}
\item The diamond substrate is first coated with three metal layers (a thick Al layer, a thin Si layer, and a thin Al layer), and a photoresist layer is spin-coated on top of this stack.
\item A soft stamp is produced by replicating the AGPM pattern from a master obtained by e-beam lithography, and used to transfer the pattern into the photoresist by solvent-assisted micro-molding. 
\item Finally, the various layers are successively etched using inductively coupled plasma reactive ion etching (ICP-RIE). By tuning the gas composition, electrical fields and pressure in the etching process, it is possible to increase the aspect ratio of the grating and obtain deep grooves in the diamond substrate.
\end{enumerate}
An etched AGPM is illustrated in Fig.~\ref{fig:agpm} as both a photograph of the complete component and a close-up of the center using a scanning electron microscope (SEM). In order to evaluate the parameters of the grating, and in particular the depth and sidewall angle (which can't be measured by SEM and other non-invasive techniques), a test sample undergoes the exact same etching process as each manufactured vortex phase mask. The test sample is cracked after etching each metal layer to monitor the parameters of the grating and to change the etching process accordingly. During the diamond etching step, the etching time is initially set to reach a too shallow final etch depth. The test sample is then cracked once more to determine the actual etch rate, allowing for a more tuned etch process that can reach the desired etch depth.

Once they have been etched, the AGPMs are evaluated in terms of coronagraphic performance. Coronagraphic testing of infrared AGPMs has mostly been carried out on the YACADIRE coronagraphic bench at Observatoire de Paris, which covers the 1--5~$\mu$m wavelength range. First results on four L-band AGPMs were reported by Delacroix et al.\cite{Delacroix13}, showing peak rejection ratios\footnote{Here, the peak rejection ratio is defined as the ratio between the intensity profiles obtained without and with the point-like source centered onto the vortex phase mask, integrated on an aperture of diameter equal to the beam width.\cite{Vargas17}} up to 500:1 for L-band operations. Eleven more L- and M-band AGPMs were produced over the last three years with our latest etching recipes and tested on the same bench, leading to broadband rejection ratios up to 1000:1\cite{Vargas17}, and thus to theoretical raw contrasts of the order of $10^{-5}$ at two beam widths from the star for a clear circular input pupil. More recently, we have obtained first results on some of these new AGPMs using the newly commissioned VODCA coronagraphic test bench at Universit\'e de Li\`ege, suggesting even higher rejection ratios, up to 2000:1 (Jolivet et al., in prep.), which is close to the theoretical broadband performance limit for such devices\cite{Vargas17}. Unfortunately, we have not been able to perform laboratory tests of our AGPMs at wavelengths longer than 5~$\mu$m so far, due to the lack of access to an appropriate mid-infrared camera. The performance of N-band AGPMs will be briefly discussed in the next section, in the context of the installation on the VLT/VISIR mid-infrared camera.


\section{COMMISSIONING AND ON-SKY OPERATIONS}
\label{sec:comm}

Based on the promising laboratory results obtained in 2012 on our L-band AGPMs\cite{Delacroix13}, it was decided to proceed with the installation of L- and N-band AGPMs into infrared imaging cameras. The first camera to receive an AGPM was VISIR, the VLT Imager and Spectrometer for the mid-InfraRed\cite{Lagage04}. The installation of the AGPM in VISIR was part of a major upgrade of the instrument, which included a new detector. The installation on VISIR included the insertion of an AGPM optimized for the 11--13.2~$\mu$m wavelength range into the VISIR focal plane wheel, as well as a new broadband filter (``12\_4\_AGP'', covering the 11.6--13.2~$\mu$m region) in the downstream pupil wheel, featuring a custom pupil stop acting as a Lyot stop. Preliminary tests carried out on the VISIR internal source suggest that the VISIR N-band AGPM provides a rejection ratio of at least 100:1. While the first on-sky results obtained with the AGPM in May 2012 were promising, VISIR was removed from the telescope due to detector issues\cite{Kerber14}. Over the last two years, VISIR has been recommissioned\cite{Kerber16}, and science verification observations have recently been performed\cite{Asmus16}. These data are currently being processed to assess the on-sky performance of the AGPM.

\begin{figure}[t]
\begin{center}
\begin{tabular}{c} 
\includegraphics[height=8cm]{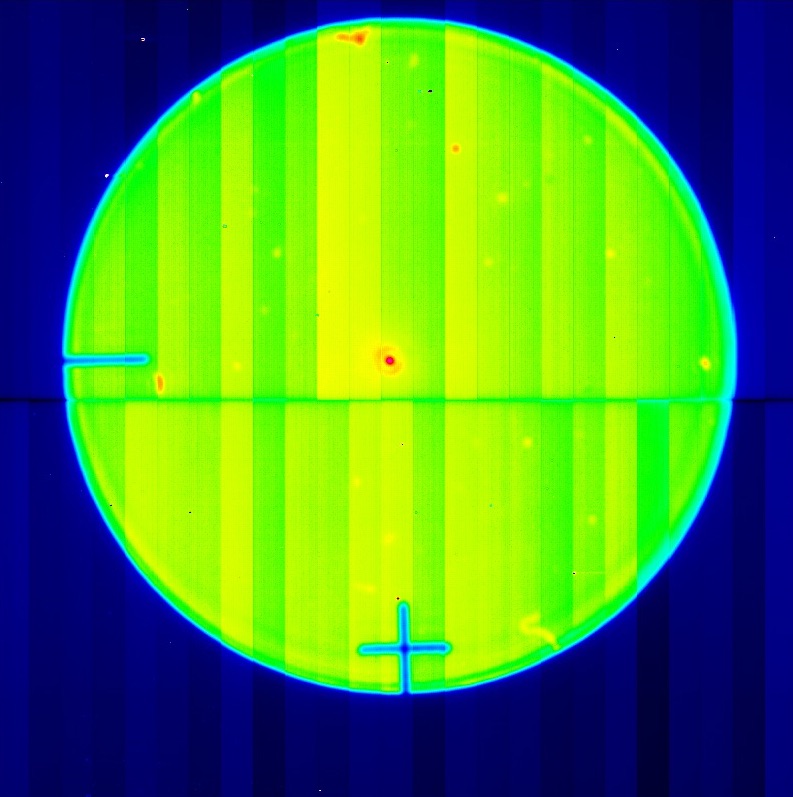} \hspace*{2mm} \includegraphics[height=8cm]{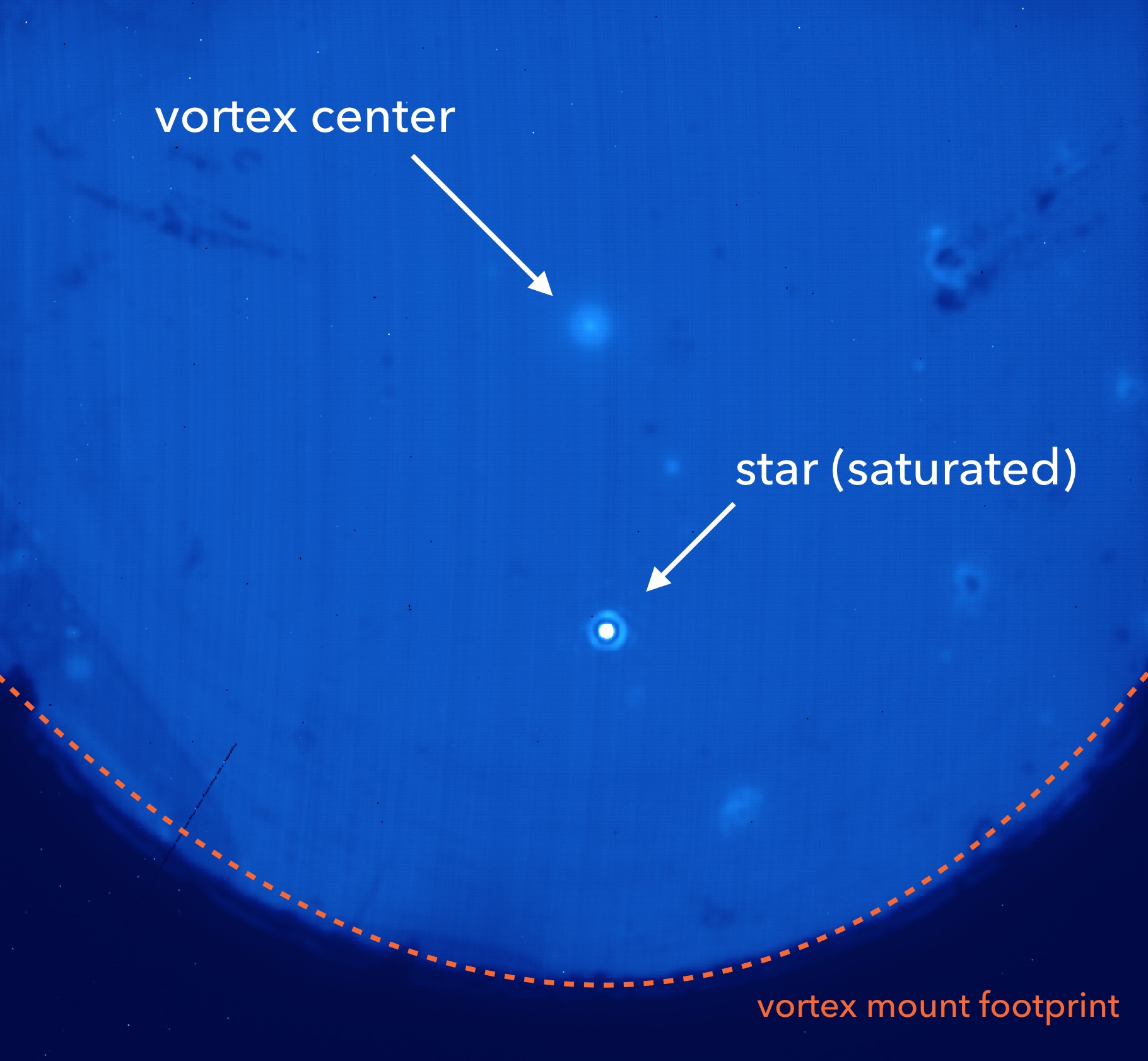}
\end{tabular}
\end{center}
\caption{\label{fig:center} \textit{Left.} Image of the VISIR N-band AGPM in front of the infrared sky, showing the dark centering marks (in blue) and the central emission peak (in red). \textit{Right.} Same for the NIRC2 L-band AGPM.}
\end{figure} 

One of the unique features of the N-band AGPM installed on VISIR is the centering marks that we have purposely written on the edge of the diamond substrate using aerosol jet printing. These centering marks, which can be easily identified through dark lines in front of a uniform background (such as the infrared sky emission, see Fig.~\ref{fig:center}), were designed to help locate the vortex center during on-sky operations. However, during the first days of commissioning, we quickly noticed that the vortex center could be clearly identified as a diffuse emission in front of the infrared sky, while its anticipated behavior was rather to create a dark hole in the background emission, as was noticed previously with optical vortices operating at shorter wavelengths. The same phenomenon was subsequently observed with the L-band AGPMs installed on three AO-assisted infrared cameras: NACO\cite{Rousset03,Lenzen03} at the VLT, LMIRCam\cite{Leisenring12} at LBT, and NIRC2 at the W.\ M.\ Keck Observatory (Fig.~\ref{fig:center}). The additional emission of the vortex above the background is generally of the order of 10\%. Based on simulations of optical propagation through a vortex coronagraph, and on dedicated lab tests, our understanding is that the bright spot is due to thermal emission coming from the warm environment around the telescope pupil (optical mounts, etc), which is partially diffracted into the geometric image of the telescope pupil by the vortex center. Although the thermal emission from the environment is not coherent, part of it can still be affected by the vortex effect, and be diffracted to another location in the pupil plane. The appearance of a bright spot at the vortex center is then explained by the fact that, in all the infrared cameras where the AGPMs have been installed so far, the AGPMs are placed upstream of the cold stop, and therefore see a significant amount of background emission coming from outside the telescope pupil. We expect that, by placing the AGPM behind a cold stop, this emission should disappear, and the vortex center emission be replaced by the textbook dark spot.\footnote{Actually, the first on-sky tests performed at LBT/LMIRCam had the vortex phase mask located downstream of a cold stop, and the emission associated with the vortex center was not observed in this configuration.} This being said, the emission from the vortex center is not a nuisance to the observations, except for a slightly increased photon noise close to the optical axis. It allows the position of the vortex center to be easily identified, removing the need for centering marks. Furthermore, it is removed (to within photon noise) by background subtraction or high-contrast imaging post-processing techniques, provided that the individual frames can be perfectly co-aligned before post-processing.

Following the installation of the N-band AGPM in VISIR, L-band AGPMs were installed in VLT/NACO in 2012\cite{Mawet13}, LBT/LMIRCam in 2013\cite{Defrere14}, and Keck/NIRC2 in 2015\cite{Femenia16}. A second AGPM was recently installed at LBT to allow observations with both 8-m apertures simultaneously. In all three cases, the installation was straightforward, as focal planes and pupil planes were already available in these instruments. The AGPM installation consisted mostly in placing the AGPM in the focal plane with a custom-made mount, which we added to an existing focal-plane mechanism. In the cases of NACO and NIRC2, the already available Lyot stops were deemed to be sufficient for vortex operations, although none of them was really optimal in terms of shape. Only the LBTI/LMIRCam camera was provided with a custom-made Lyot stop for vortex operations. In all cases, the commissioning of the new vortex mode was performed within only a couple of nights, as it mostly followed a standard observing template of the instrument (only LMIRCam was not previously equipped with a focal-plane coronagraph before the AGPM installation).

The most challenging aspect of on-sky vortex operations is the accurate centering of the star onto the vortex mask. Not only does the star need to be centered within a small fraction of the diffracted beam width, but this alignment must also be preserved during the whole observing sequence. To this end, we have adapted an alignment method based on the science image, originally proposed for the four-quadrant phase mask\cite{Mas12}. This alignment method, referred to as QACITS\cite{Huby15}, has now been tested on the three L-band cameras discussed here. The initial validation and the most thorough tests were carried out at Keck/NIRC2, thanks to the allocation of engineering time by the Caltech Optical Observatories. This engineering time was instrumental to the on-sky validation of the QACITS algorithm, which is now available in shared-risk mode for the general observer at Keck/NIRC2\cite{Huby16} and will soon be fully integrated into the NIRC2 control software. The main features of the QACITS algorithm are the following:
\begin{itemize}
\item fully automatic procedure to center the star onto the vortex mask, including all required calibration measurements (sky, PSF);
\item closed-loop operation with adjustable loop gain and set point during scientific acquisitions;
\item closed-loop pointing accuracy down to about $0.02\lambda/D$ rms (i.e., 2~mas at Keck), ensuring high coronagraphic rejection and consistent data quality from star to star and from night to night.
\end{itemize}
Based on this successful commissioning at Keck/NIRC2, we expect to deploy QACITS on the other infrared cameras in the coming months.


\section{ON-SKY CORONAGRAPHIC PERFORMANCE}
\label{sec:perfo}

While in an ideal case (i.e., for a perfect wavefront), the diffraction pattern associated with the starlight is expected to be constant in time and therefore to cancel out during post-processing, reducing the amount of starlight in the focal plane of the telescope is still a critical step to reach high contrast, for two main reasons. First, the photon noise associated to the diffracted starlight can completely overwhelm the signal from a faint planetary companion. Second, due to time-varying optical aberrations (related to atmospheric turbulence and to instrumental imperfections), the diffraction pattern becomes variable in time. One of the main complications to this variability is that the speckles created by the residual atmospheric turbulence and optical aberrations can interfere with the diffraction pattern associated to the star, leading to bright, semi-static ``pinned'' speckles.\cite{Bloemhof04,Soummer07} Reducing the amount of diffracted starlight greatly helps in mitigating this effect.

Four main effects contribute to the imperfect cancellation of starlight by the vortex coronagraph during observations (Serabyn et al., in prep): (i) diffraction due to the central obstruction of the telescope and to other deviations from a clear circular aperture, (ii) pointing errors, (iii) wavefront errors other than pointing, and (iv) imperfections in the vortex phase mask. These four contributions vary from one instrument to another. Typical orders of magnitude can nevertheless be given for the AO-assisted L-band cameras. 
\begin{itemize}
\item First, the leakage (i.e., fraction of stellar energy reaching the plane of the detector) due to the central obstruction scales as $a^2/R^2$, where $a$ is the radius of the central obstruction and $R$ the pupil radius\cite{Jenkins08}. A typical obstruction $a/R=0.2$ gives a leakage of 0.04. This contribution is constant in time and therefore cancels out during post-processing.
\item Second, the additional leakage due to a pointing error $\theta$ has been derived in the Appendix of Huby et al.\cite{Huby15}, and is given by $(\pi\theta D/\lambda)^2/8$. Assuming that low-frequency pointing drifts are corrected by QACITS, we are mostly faced with high-frequency residual tip-tilt jitter (either due to atmospheric turbulence or telescope vibrations). Assuming a typical residual tip-tilt jitter of 10~mas rms for standard adaptive optics systems, this leakage term is of the order of 0.01. While this value is smaller than the diffraction-induced leakage discussed above, it can still represent the limiting factor for the reachable contrast after post-processing, because this contribution is variable in time and generally does not cancel out during post-processing. In the case of VLT/VISIR, the lack of adaptive optics results in a large pointing jitter, of the order of 100~mas rms or more, which leads to a leakage larger than 0.1.
\item Third, assuming a good AO wavefront correction level, the additional leakage due to higher-order aberrations (beyond tip-tilt) can be approximated by $1-S$, where $S$ is the Strehl ratio. At L band, the Strehl ratio is generally higher than 80\% for the AO-assisted cameras considered here, and can reach up to 95\% in LMIRCam thanks to the LBTAO system\cite{Skemer14}. The order of magnitude of this contribution is therefore around 0.1, while its standard deviation is roughly of the order of 0.01 under stable atmospheric conditions with good AO correction (i.e., variations of Strehl ratio around 1\% during the observations).
\item Finally, the intrinsic chromatic leakage due to the imperfections of the vortex phase mask has been evaluated in the lab, and is always smaller than 0.01 for the AGPMs that we used on sky. This contribution is thus always negligible, and is removed during post-processing anyway, as it is constant in time.
\end{itemize}

\begin{figure}[p]
\begin{center}
\includegraphics[height=8.5cm]{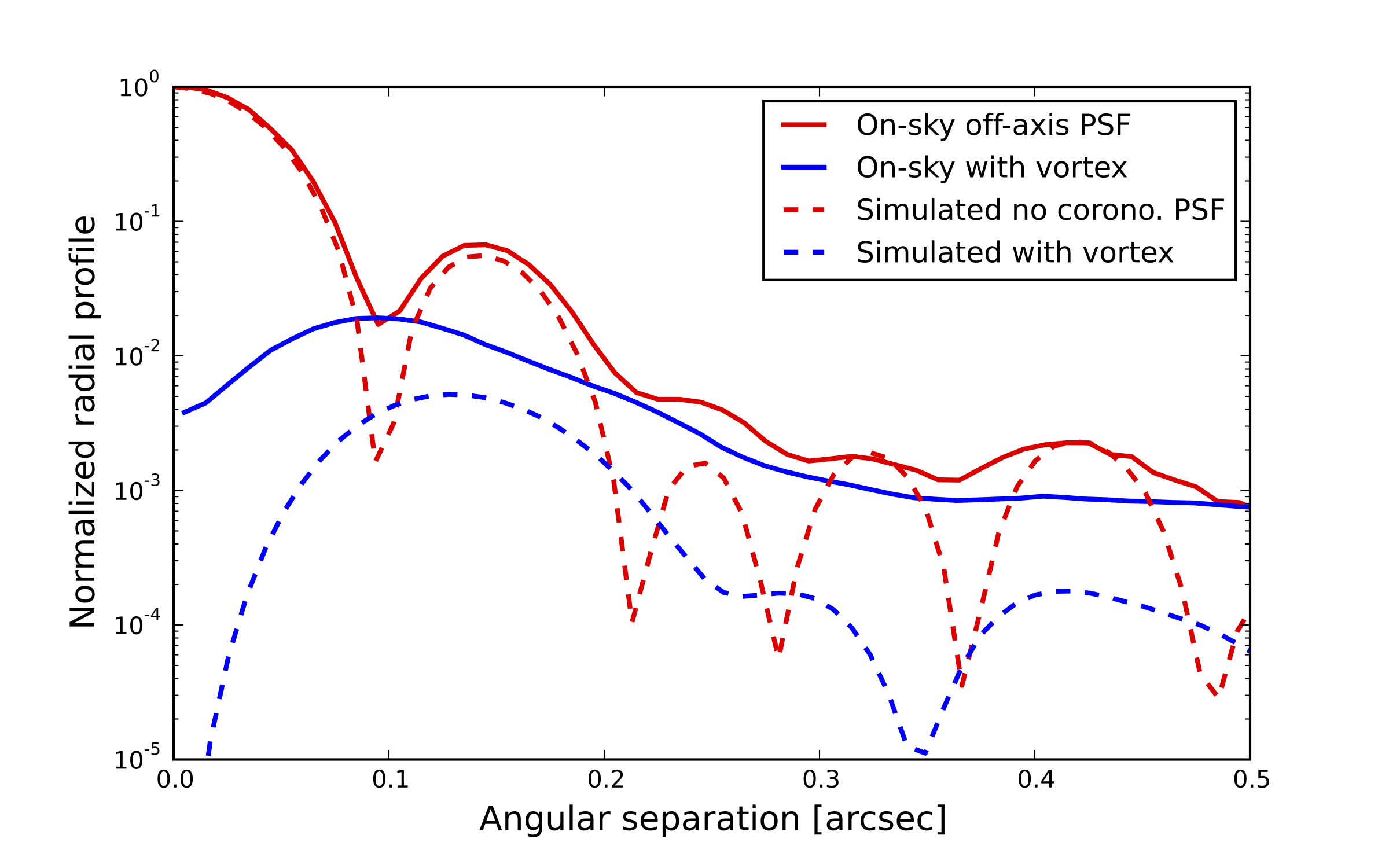}
\end{center}
\caption{\label{fig:profile} Typical intensity profiles with the star centered (blue solid line) or not (red solid line) onto the vortex phase mask at Keck/NIRC2. Simulated intensity profiles computed for the Keck pupil with and without vortex coronagraph assuming a perfect input wavefront are shown in dashed lines with the same color code for comparison.}
\end{figure} 

In Fig.~\ref{fig:profile}, we show the radial profiles of the NIRC2 PSF with the star aligned or not onto the vortex phase mask center (solid lines). While the central part of the PSF is largely cancelled by the vortex, the gain in terms of brightness in the wings is much weaker. This behavior is expected in the presence of optical aberrations, which create an incoherent halo that cannot be cancelled out by the vortex coronagraph. The stellar light diffracted by the central obstruction also contributes to the off-axis intensity profile, as demonstrated by the dashed lines in Fig.~\ref{fig:profile}, which have been obtained by simulations of the optical propagation through the NIRC2 instrument taking into account the Keck pupil and assuming an ideal input wavefront. Even though the wings are not largely attenuated, providing a strong cancellation of the stellar peak (within $1\lambda/D$ from the center) still has two immediate benefits: (i) it enables to increase the integration time without reaching saturation, and (ii) it reveals low-brightness features (speckles and planets) that were otherwise outshone by the bright stellar peak. These two benefits were clearly illustrated during the science verification of the AGPM on VLT/NACO, where observations of the bright star $\beta$~Pictoris revealed its faint planetary companion in the raw frames and in the real-time display while observing at the telescope, without any post-processing\cite{Absil13}. Getting a better view of the speckles in the raw frames also improves our capability to sense and correct them. This is now exploited at Keck/NIRC2, where a speckle nulling algorithm has been deployed, and is used to reduce the level of optical aberration in the instrument during daytime\cite{Bottom16}.

\begin{figure}[p]
\begin{center}
\begin{tabular}{cc} 
& \vspace*{-4cm} \includegraphics[height=4cm]{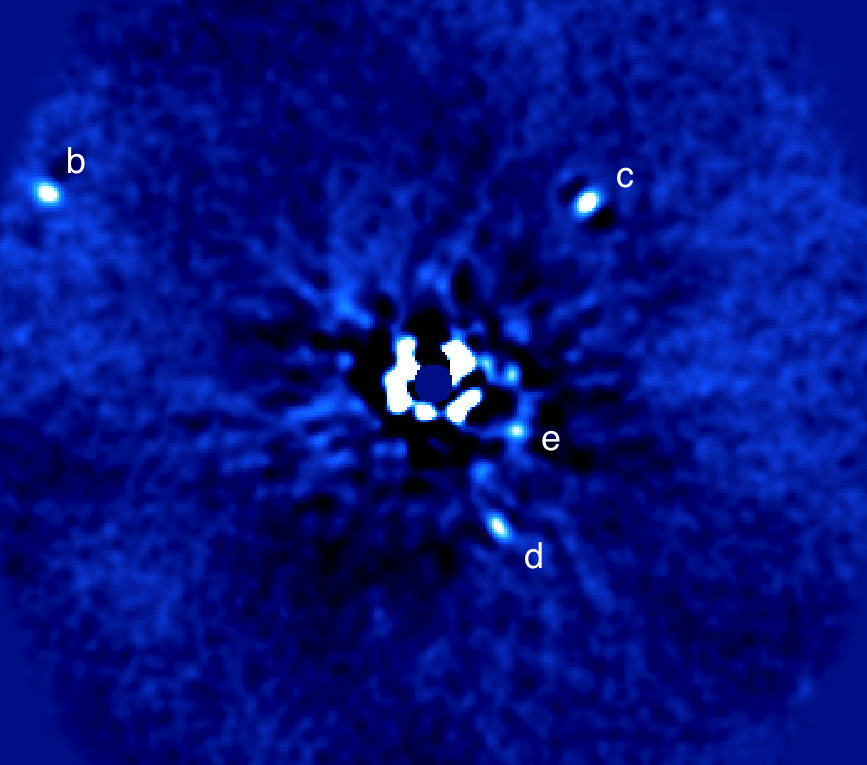} \\
\includegraphics[height=8.5cm]{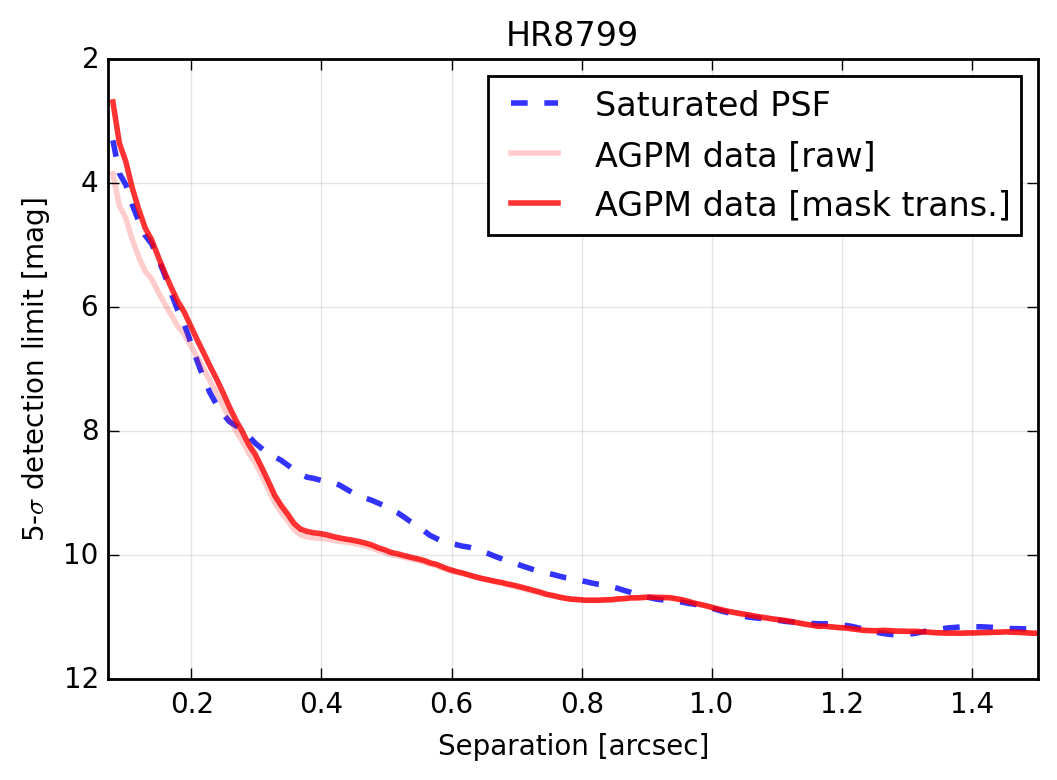} & \includegraphics[height=4cm]{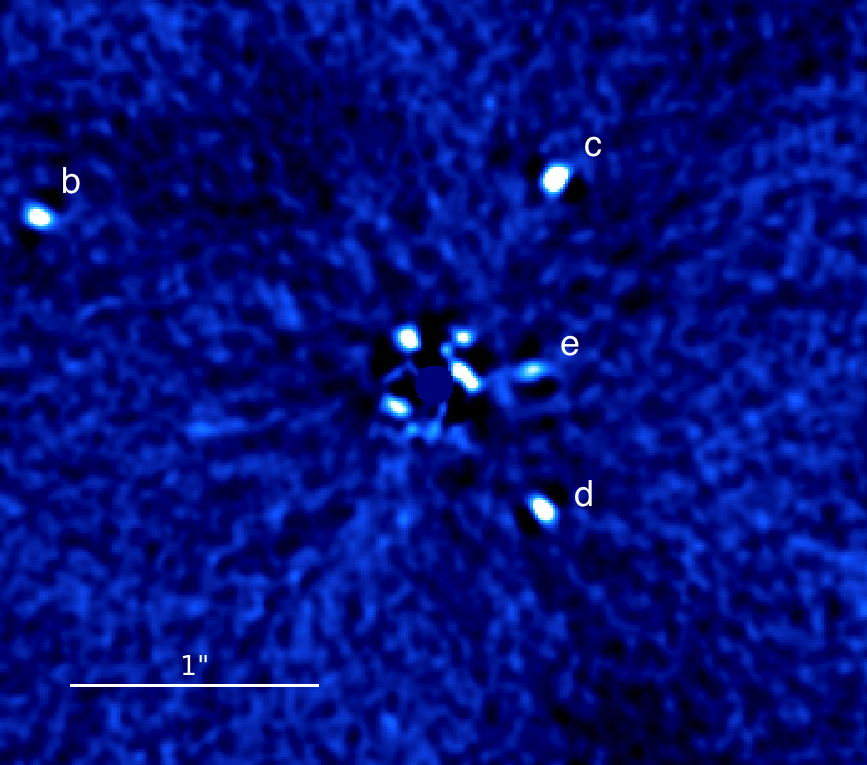}
\end{tabular}
\end{center}
\caption{\label{fig:limdet} \textit{Left.} Sensitivity limits in terms of planet/star contrast around HR8799, based on PCA-ADI post-processing of data obtained at Keck/NIRC2 with saturated imaging (blue dotted line) and with vortex coronagraphy imaging enhanced with speckle nulling (red solid line). A thin red line also shows the raw sensitivity limit, which does not take into account the intrinsic off-axis transmission of the vortex coronagraph. \textit{Right.} Illustration of the final image obtained after PCA-ADI post-processing of these two data sets (top: saturated imaging -- bottom: vortex coronagraphy imaging).}
\end{figure} 

The next step is to evaluate how the vortex phase mask improves the performance of the considered instruments for faint companion detection. Based on the intensity profiles shown in Fig.~\ref{fig:profile} for Keck/NIRC2, we do not expect a large gain in terms of detection limits, because the starlight rejection achieved by the vortex coronagraph is severely limited by optical aberrations. To assess the performance of the new vortex mode at Keck/NIRC2 and compare it to the standard saturated imaging mode for exoplanet detection, we reduced with the same post-processing algorithm two data sets obtained on the bright star HR8799 under similar atmospheric conditions, for a similar integration time, and with a similar parallactic angle rotation range. The saturated imaging data set was obtained on 2009-11-01. It consists of 20 min of total on-source time with 202$^{\circ}$ of parallactic angle rotation. Another data set was obtained on 2015-10-24 with the vortex coronagraph. A speckle nulling solution computed during daytime on the internal source was applied on sky during that night (it was the first night speckle nulling was tested at Keck). This data set has 30~min of on-source time, for a parallactic angle rotation of 190$^{\circ}$. The two data sets were processed using the full-frame PCA-ADI algorithm included in the VIP python package\cite{Gomez16}. The number of principal components used to model and subtract the PSF in each frame was tuned to get the best possible signal-to-noise ratio on HR8799e, resulting in seven principal components for the saturated imaging data set, and sixteen components in the vortex imaging data set. The central $1\lambda/D$ region was masked out during post-processing in both cases, to prevent the PCA-ADI algorithm from trying to reproduce a scientifically uninteresting region. The sensitivity limits in terms of planet/star contrast (aka contrast curves) are plotted in Fig.~\ref{fig:limdet}. They have been computed using the prescription of Mawet et al.\cite{Mawet14}, where the noise estimation is based on the variance of the flux integrated on independent $\lambda/D$-sized apertures at a given angular separation, including the penalty for small sample statistics. The algorithm throughput is taken into account in our sensitivity limits, through the injection and recovery of fake companions into the raw data. In order not to bias the sensitivity limits, the four planets were first subtracted from the raw data cube using the negative fake companion technique\cite{Marois10,Lagrange10}. The vortex imaging data set shows a significant improvement upon the saturated imaging data set in terms of sensitivity in the 3 to $10\lambda/D$ region (i.e., $0.25''$ to $0.8''$ for Keck/NIRC2 operating at L band). The final images obtained after PCA-ADI post-processing on the two original data sets (with planets) is also shown in Fig.~\ref{fig:limdet}. Visual inspection shows largely reduced residuals in the innermost regions, where planets HR8799e and HR8799d are located.


\section{FIRST SCIENTIFIC RESULTS}
\label{sec:science}

While direct imaging has led to a handful of detections of giant planets orbiting at several tens of AU from their parent star, arguably the most important result of direct planet imaging so far has been to constrain the occurrence rate of giant planets at large orbital distance from their host star. To do so, planet population analyses have relied on contrast curves, which can be translated into mass sensitivity limits based on planet evolutionary models. One of the first applications of the newly commissioned vortex modes has therefore been to improve upon state-of-the-art sensitivity limits in terms of mass, and in particular to revisit known extrasolar planetary systems to look for additional planets that might have remained unnoticed by other observing modes until now. This is what we did on HR8799 with LBT/LMIRCam and Keck/NIRC2, as well as on $\beta$~Pic with VLT/NACO. The HR8799 observations with LBT/LMIRCam have already been reported elsewhere\cite{Defrere14}, and the $\beta$~Pic data set as well\cite{Absil13}. Here, we plot the sensivity limits obtained with the three instruments on these bright stars in the same figure to give a (partially unfair) impression of how these instruments compare in terms of performance (see Fig.~\ref{fig:compar}). The relevant information needed to compare these sensitivity limits is given below:
\begin{itemize}
\item \textbf{VLT/NACO}. A total on-source time of 1h54 was obtained on the bright star $\beta$~Pic ($L=3.4$). The total sequence lasted for more than 3h, resulting in a total parallactic angle rotation of 83$^{\circ}$.
\item \textbf{LBT/LMIRCam}. A total on-source time of 1h20 was obtained on HR8799 ($L=5.2$). The total sequence lasted for more than 2h, resulting in a parallactic angle rotation of 96$^{\circ}$.
\item \textbf{Keck/NIRC2}. A total on-source time of 30min was obtained on HR8799 ($L=5.2$). The total sequence lasted for more than 1h, resulting in a parallactic angle rotation of 168$^{\circ}$.
\end{itemize}

\begin{figure}[p]
\begin{center}
\includegraphics[height=8.5cm]{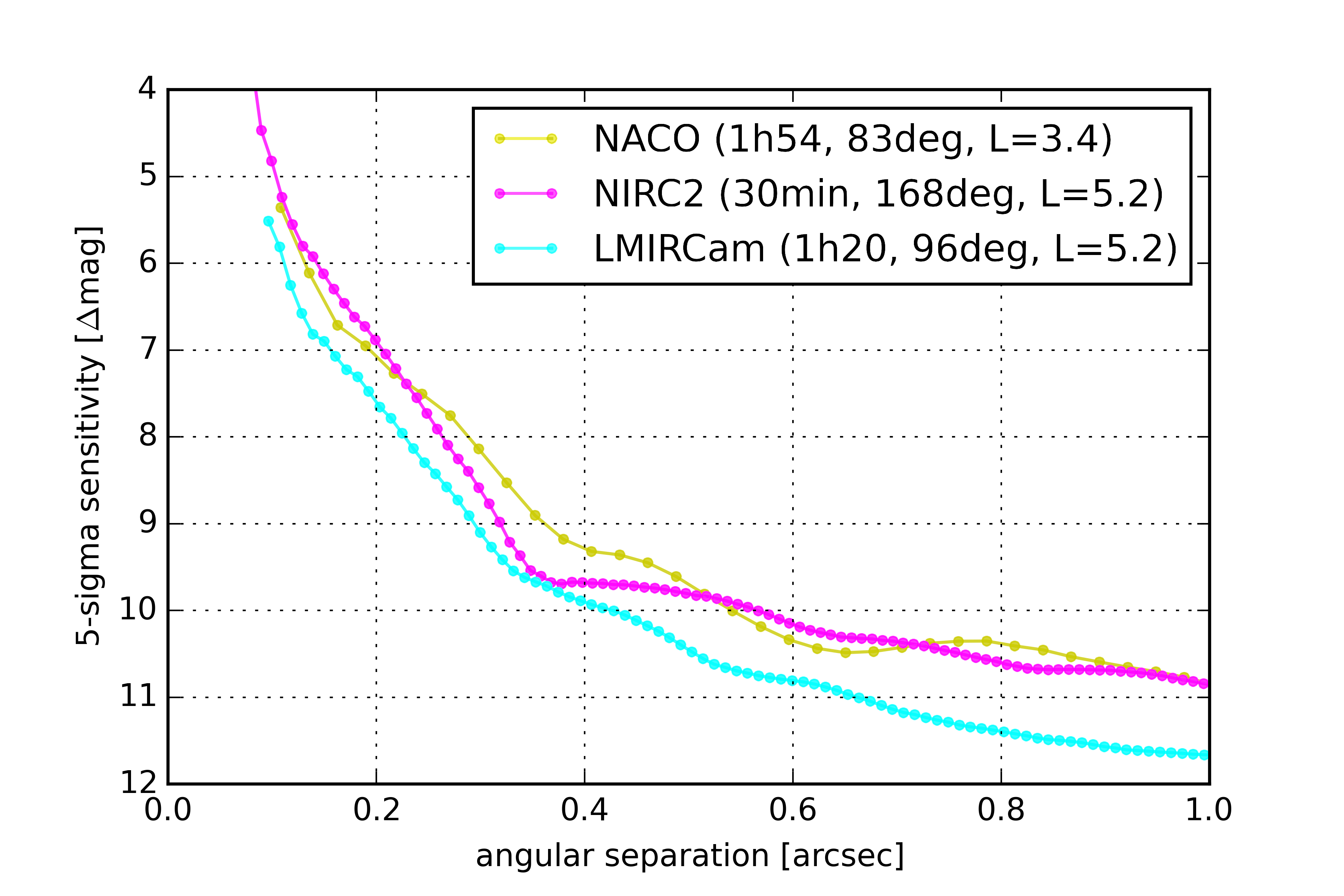}
\end{center}
\caption{\label{fig:compar} Comparison of the sensitivity limits in terms of planet/star contrast for three data sets obtained with the thermal infrared vortex coronagraph on three different instruments. See text for details.}
\end{figure} 

Based on Fig.~\ref{fig:compar}, we conclude that the low background and high-quality adaptive optics correction provided by the LBT to LMIRCam is the best combination to reach deep contrasts. The fact that the Lyot stop was optimized for the vortex coronagraph in LMIRCam also helps reach lower starlight residuals. The contrasts achieved by NACO and NIRC2 are similar, although the background limit seems to be a couple magnitudes deeper for NIRC2 based on the magnitude of the target star and on the integration time used in these observations. Making more detailed conclusions based on just three data sets is not possible, but this plot certainly shows the merits of a careful instrument design for high-contrast applications in the thermal infrared.

\begin{figure}[p]
\begin{center}
\includegraphics[height=8.5cm]{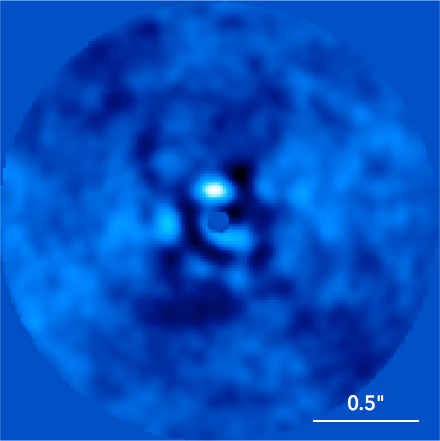}
\end{center}
\caption{\label{fig:hd169142} Detection of a mysterious point-like source at 1.5 beam width from the pre-main sequence star HD169142 using the AGPM vortex coronagraph at VLT/NACO (adapted from Reggiani et al.\cite{Reggiani14}).}
\end{figure} 

A specific feature of the vortex coronagraph is to enable the detection of companions down to an inner working angle of one beam width ($\lambda/D$). This capability has already been demonstrated with a vortex phase mask operating at shorter wavelengths on the well-corrected subaperture at Palomar\cite{Serabyn10,Mawet11}. The small-angle tradition continues with the AGPM operating at longer wavelengths, in particular with the discovery of a candidate protoplanet around the Herbig Ae star HD169142 (Fig.~\ref{fig:hd169142}). The detection of this candidate companion, reported by two different teams using the vortex coronagraph of VLT/NACO\cite{Reggiani14,Biller14}, has been enabled by the combination of L-band observations with high contrast performance at very small angular separation ($0.15''$ in this case, which represents $1.5\lambda/D$ for NACO). Follow-up observations at shorter wavelengths have failed to confirm this detection until now, which could point towards an accreting protoplanet associated with significant accretion luminosity at L band. The capability of the vortex coronagraph to access small separations in the thermal infrared is now used to search for companions around young stars. It also makes the thermal infrared vortex a natural complement to recently commissioned planet imaging instruments, which access these very small angular separations at shorter wavelengths. The results of these observing campaigns will be presented in future publications.


\section{LESSONS LEARNED AND PERSPECTIVES}
\label{sec:lessons}

One of the main lessons learned from the commissioning and scientific exploitation of AGPMs on world-leading infrared cameras is that reaching deep contrasts requires more than just inserting a phase mask in the focal plane of the instrument. First, it is equally important to include an optimal Lyot stop, which is not always the case for the instruments described here (e.g., NACO, NIRC2). Ideally, the vortex coronagraph should also be paired with an optimized upstream apodizer, to reduce the effect of the central obstruction on the residual starlight level\cite{Mawet13b}. In any case, for the general-purpose instruments described here (as opposed to instruments dedicated to high-contrast imaging such as VLT/SPHERE or Gemini/GPI), the on-sky performance of the vortex coronagraph is generally limited by the level of optical aberrations seen by the vortex phase mask, either due to residual atmospheric turbulence or to non-common path aberrations in the instrument between the AO wavefront sensor and the coronagraph. It must be kept in mind that the vortex coronagraph only acts on the coherent part of the starlight, and that most of the incoherent halo remains unaffected downstream of the coronagraph. A global optimization of the instrument needs to be considered to reach the highest possible coronagraphic performance. Such an optimization is currently underway in the context of the METIS instrument\cite{Brandl16} for the E-ELT, where high contrast imaging is one of the main drivers for the instrument design.\cite{Kenworthy16} Another major lesson learned is that operating the vortex coronagraph without an automatic centering and pointing control algorithm is a demanding task that generally leads to inconsistent data quality (depending mostly on the person in charge of the vortex operations). And of course, we once again realized that optimal data processing is also a key part of high-contrast observations. This is why we developed an open-source image processing package called VIP\cite{Gomez16}, which includes innovative algorithms for companion detection in high-contrast data sets\cite{Gomez16b}.

Although installing a vortex phase mask in an existing, coronagraphy-enabled instrument does not solve all the high-contrast imaging challenges at once, we can still note some immediate benefits from the presence of the vortex phase mask (and associated Lyot stop). First, the vortex coronagraph significantly reduces the amount of starlight in the camera, thereby reducing photon noise and enabling longer integration times. By reducing the strength of the starlight diffraction pattern, the vortex coronagraph also reduces the brightness of pinned speckles, which result from the constructive interferences between the diffraction pattern and aberration-induced speckles. The vortex also helps reveal instrumental imperfections in general, which in turns facilitates the implementation of corrective actions (such as speckle nulling). Furthermore, the on-sky operations of the vortex coronagraph benefit from a well-tested framework, including automatic centering and close-loop pointing control during scientific measurements, which results in consistent, high data quality.

Although the vortex coronagraph has already shown very promising on-sky results, there is still a lot of room for improvement in the coming years. Among the perspectives for the vortex coronagraph, we are planning to focus on four main topics:
\begin{itemize}
\item \textbf{Shorter wavelengths}. So far, our vortex phase masks based on sub-wavelength gratings have addressed wavelengths longer than 3~$\mu$m. By reducing the period of the grating, we are planning to address shorter wavelengths (even though near-infrared wavelengths are already addressed by vortex phase masks made up of liquid crystal polymers\cite{Mawet09}). This requires controlling the etching process even better, as errors on the grating parameters will have a larger impact at shorter wavelengths. First results obtained at K band are promising, and could enable the installation of such vortex masks in XAO-assisted near-infrared high contrast imagers in the near future.
\item \textbf{Higher topological charges.} The topological charge $l$ of the vortex phase mask is the number of times the phase wraps from 0 to $2\pi$ in one circuit about the center. Only vortices featuring even topological charges produce a coronagraphic effect, and the off-axis transmission close to the optical axis scales as $\theta^l$, with $\theta$ the angular separation to the optical axis. Higher topological charges therefore result in a broader extinction pattern close to the optical axis, and thus in a reduced sensitivity to pointing jitter\footnote{this is especially important in the presence of high-frequency pointing jitter, due e.g.\ to telescope vibrations that are generally poorly corrected by AO systems and too fast to be sensed by the QACITS pointing control algorithm} and source spatial extension, at the expense of a degraded inner working angle. We are currently investigating the design and manufacturing of charge-4 vortices based on sub-wavelength gratings\cite{Delacroix16}.
\item \textbf{Optimal apodization.} It has been shown that a ring apodizer placed in an upstream pupil plane can mitigate the effect of diffraction by the central obstruction of the telescope, at the expense of a reduced throughput.\cite{Mawet13b} More recently, it was proposed to use an apodizing phase mask in conjunction with the Lyot stop in the downstream pupil plane to create a deep dark hole in the final focal plane, at the expense of a reduced encircled energy in a $1\lambda/D$ aperture for the off-axis companion PSF.\cite{Ruane15} These two concepts can both enable improved sensitivity to off-axis companions. However, their optimization needs to take into account the constraints of ground-based observations (i.e., speckle noise and thermal background emission), which generally require more weight to be put on the coronagraphic throughput in the optimization process. We are in the process of investigating the combination of these concepts, as well as optimized binary amplitude Lyot stops, to reach the best possible ground-based planet detection performance.\cite{Carlomagno16,Ruane16}
\item \textbf{Low-order wavefront sensing.} Finally, we intend to explore and test in the laboratory different ways to combine the vortex coronagraph with low-order wavefront sensing (LOWFS). We are particularly focusing on LOWFS solutions based on the science image, because they are sensitive to all aberrations affecting the coronagraph, including non-common path aberrations.\cite{Singh14}
\end{itemize}

\acknowledgments 
 
We thank P.\ Baudoz and J.\ Parisot for maintaining and sharing with us the YACADIRE coronagraphic test bench at Observatoire de Paris-Meudon. We also thank all the people involved in the installation, commissioning, and operations of the AGPM at VLT, LBT, and Keck. The research leading to these results has received funding from the European Research Council under the European Union's Seventh Framework Programme (ERC Grant Agreement n.\ 337569), and from the French Community of Belgium through an ARC grant for Concerted Research Action. The W. M. Keck Observatory is operated as a scientific partnership among the California Institute of Technology, the University of California, and the National Aeronautics and Space Administration. The Observatory was made possible by the generous financial support of the W.\ M.\ Keck Foundation. Part of this work was carried out at the Jet Propulsion Laboratory, California Institute of Technology, under contract with NASA. The LBTI is funded by the National Aeronautics and Space Administration as part of its Exoplanet Exploration Program. The LBT is an international collaboration among institutions in the United States, Italy and Germany. LBT Corporation partners are: The University of Arizona on behalf of the Arizona university system; Instituto Nazionale di Astrofisica, Italy; LBT Beteiligungsgesellschaft, Germany, representing the Max-Planck Society, the Astrophysical Institute Potsdam, and Heidelberg University; The Ohio State University, and The Research Corporation, on behalf of The University of Notre Dame, University of Minnesota and University of Virginia. This research was supported by NASA's Origins of Solar Systems Program, grant NNX13AJ17G.

\bibliography{thermal_ir_vvc} 
\bibliographystyle{spiebib} 

\end{document}